\documentclass[traditabstract]{aa}
\usepackage{txfonts}
\usepackage{graphicx}
\usepackage{natbib}

\bibpunct{(}{)}{;}{a}{}{,}

\newcommand{\oam}{\mbox{\object{AM\,Her}}}
\newcommand{\amher}{AM\,Her}

\newcommand{\ten}[2]{#1\times 10^{#2}}

\newcommand{\msunyr}{M$_\odot\,\mathrm{yr}^{-1}$}
\newcommand{\msun}{M$_\odot$}

\newcommand{\nh}{$N_\mathrm{H}$}

\newcommand{\atoms}{H-atoms\,cm$^{-2}$}
\newcommand{\xh}{$x_\mathrm{H}$}
\newcommand{\xhe}{$x_\mathrm{He}$}

\newcommand{\erg}{erg\,s$^{-1}$}
\newcommand{\ergs}{erg\,cm$^{-2}$s$^{-1}$}
\newcommand{\ergsa}{erg\,cm$^{-2}$s$^{-1}$\AA$^{-1}$}

\newcommand{\lya}{Ly$\alpha$}

\newcommand{\lacc}{$L_{\rm acc}$}

\newcommand{\mdot}{${\dot M}$}

\newcommand{\fbb}{$F_{\rm bb}$}

\newcommand{\rosat}{\textit{ROSAT}}
\newcommand{\chandra}{\textit{Chandra}}
\newcommand{\euve}{\textit{EUVE}}

\begin{document}

\title{A new soft X-ray spectral model for polars with an application\\
  to AM Herculis
\thanks{Based on observations with the \chandra\ and \rosat\ satellites.}
}

\author {K.\ Beuermann \inst{1}
\and     V. Burwitz{\inst{2}}     
\and     K. Reinsch \inst{1}}

%  offprint comment removed because we do not distribute reprints anymore !

\institute{Institut f\"ur Astrophysik, Friedrich-Hund-Platz 1,
D-37077 G\"ottingen, Germany
%, beuermann or reinsch@astro.physik.uni-goettingen.de,
%reinsch@astro.physik.uni-goettingen.de
\and
Max-Planck-Institut f\"ur Extraterrestrische Physik, Garching b. M\"unchen
%, burwitz@mpe.mpg.de
}

\date{Received 14 March 2012/  accepted \ldots 2012}
 
\authorrunning{K. Beuermann et al.}  
\titlerunning{A new soft X-ray spectral model for polars}

\abstract{ We present a simple heuristic model for the time-averaged
  soft X-ray temperature distribution in the accretion spot on the
  white dwarf in polars. The model is based on the analysis of the
  \chandra\ LETG spectrum of the prototype polar AM Her and involves
  an exponential distribution of the emitting area vs. blackbody
  temperature $a(T)=a_0\,\mathrm{exp}(-T/T_0)$. With one free
  parameter besides the normalization, it is mathematically as simple
  as the single blackbody, but is physically more plausible and fits
  the soft X-ray and far-ultraviolet spectral fluxes much better.  The
  model yields more reliable values of the wavelength-integrated flux
  of the soft X-ray component and the implied accretion rate than
  reported previously.  }


\maketitle

\section{Introduction}

In the 'polar' subtype of cataclysmic binaries, a magnetic white-dwarf
mass accretes matter from its Roche-lobe filling main-sequence
companion. Part of the energy released in the process appears as soft
X-ray emission of the heated photosphere in the accretion spot(s) near
the magnetic pole(s) of the white dwarf. The emitted spectrum is
conventionally modeled as a single blackbody with a temperature around
30\,eV \citep[e.g.][]{ramsayetal96,ramsaycropper04}, although on
theoretical grounds the source is not expected to be
isothermal. Constructing a viable model requires observations that
extend from the soft X-ray into the far-ultraviolet regime with
sufficient spectral resolution and good counting statistics,
preferably from a source that exhibits a single accretion spot. A
suitable source is the prototype polar \oam\ in its high state, when
it is the brightest soft X-ray source in the sky. The first
high-resolution soft X-ray spectrum of \amher\ was taken by
\citet{paerelsetal96} with the \euve\ short-wavelength
spectrometer. Because of the limited spectral coverage, however, they
retained the single-blackbody assumption.  Substantially improved
high-resolution soft X-ray spectra of \amher\ became available with
the Low Energy Transmission Grating Spectrometer (LETGS) on board of
\chandra\ \citep{burwitzetal02,trill06} with the High Resolution
Camera for Spectroscopy (HRC-S) as detector. In this paper, we analyze
the orbital mean LETG spectrum of \amher\ taken in September 2000.

Intermittent heating and cooling determines the mean temperature
distribution averaged over time and the spatial coordinates in the
spot. The distribution ranges from the photospheric temperature to
several tens of eV \citep[see][for the case of \amher]{beuermannetal08}. 
We determined the mean temperature distribution for \amher\ by
unfolding the time-averaged \chandra\ LETG spectrum of \amher. Our
model employs a quasi-continuous distribution of blackbodies (BB) that
is described by a single free parameter besides the common
normalization. Mathematically, this model is as simple as the single
blackbody. We fitted the LETG spectrum of \amher\ by adding the
thermal emission of the post-shock cooling flow to the soft X-ray
component. A similar model was adopted by \citep{traulsenetal10} for
the XMM Newton spectrum of the polar AI\,Tri.

\section{Observations of \amher\ and data analysis}

\begin{figure*}[t]
\includegraphics[height=97mm,bb=110 58 544 621,angle=270,clip]{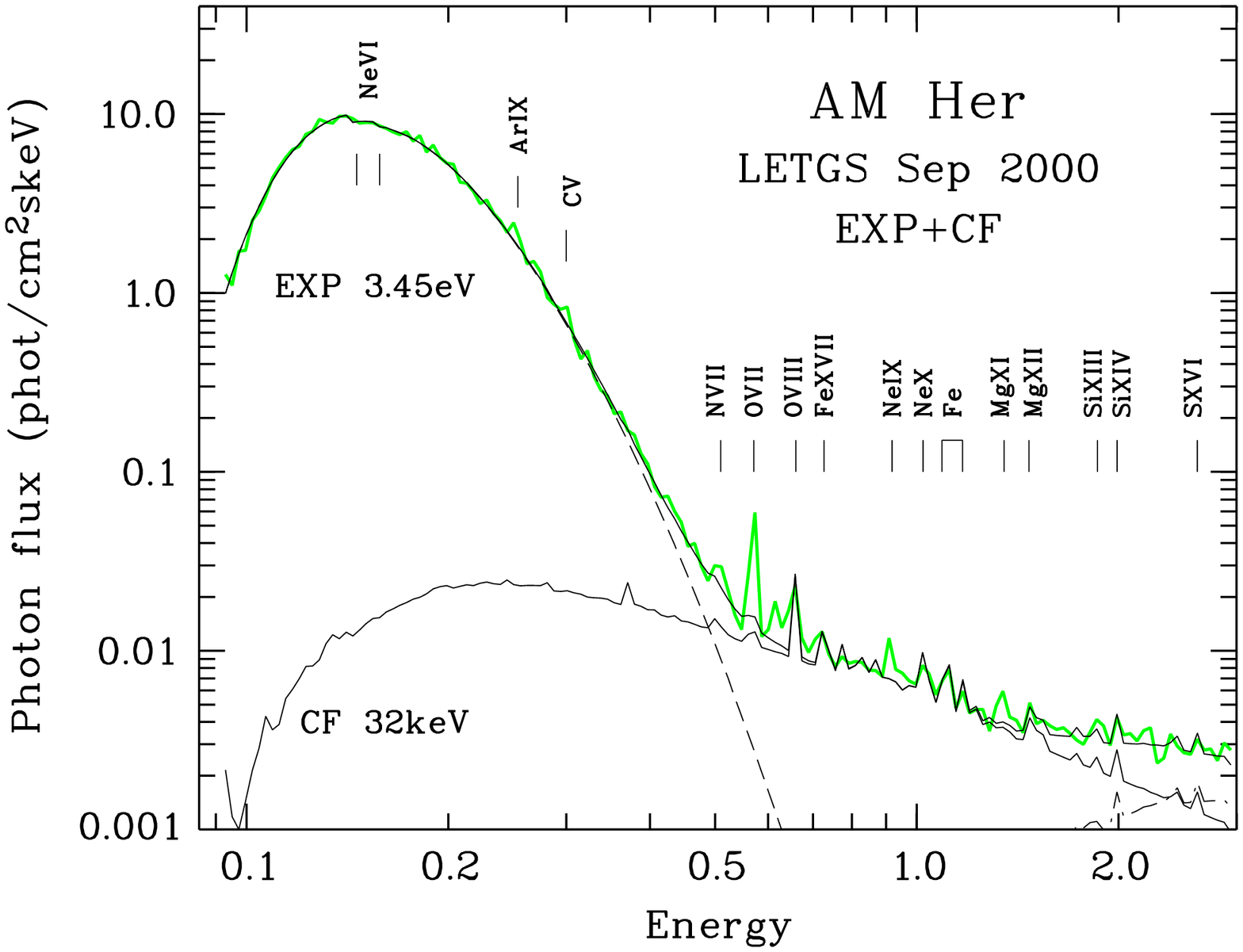}
\hfill
\begin{minipage}[t]{77mm}
\vspace{0.0mm}
\includegraphics[width=76.6mm,angle=0,clip]{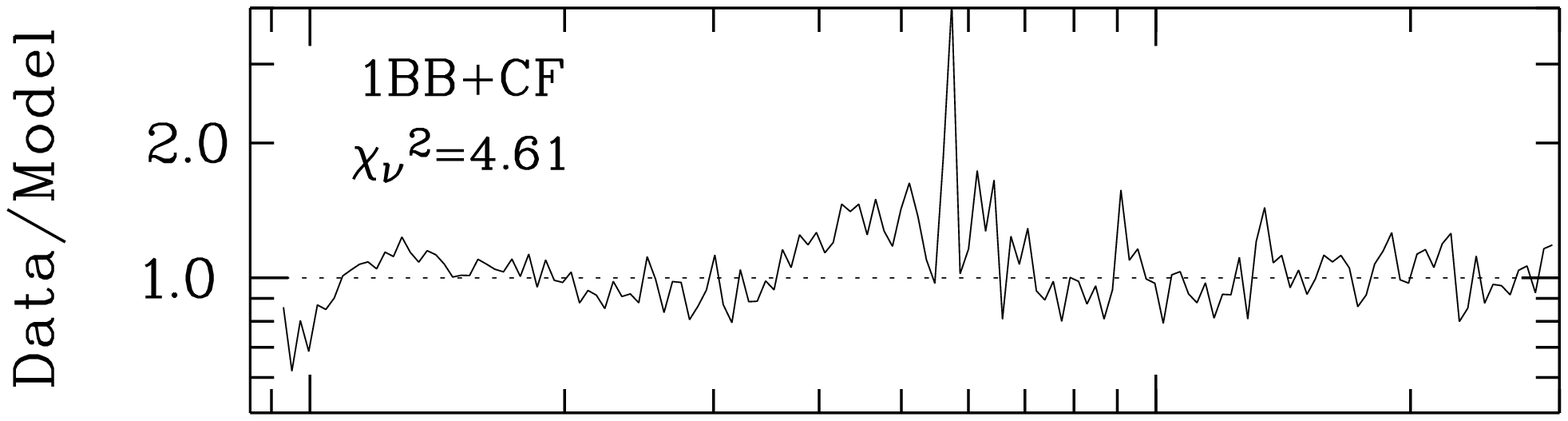}
\includegraphics[width=76.6mm,angle=0,clip]{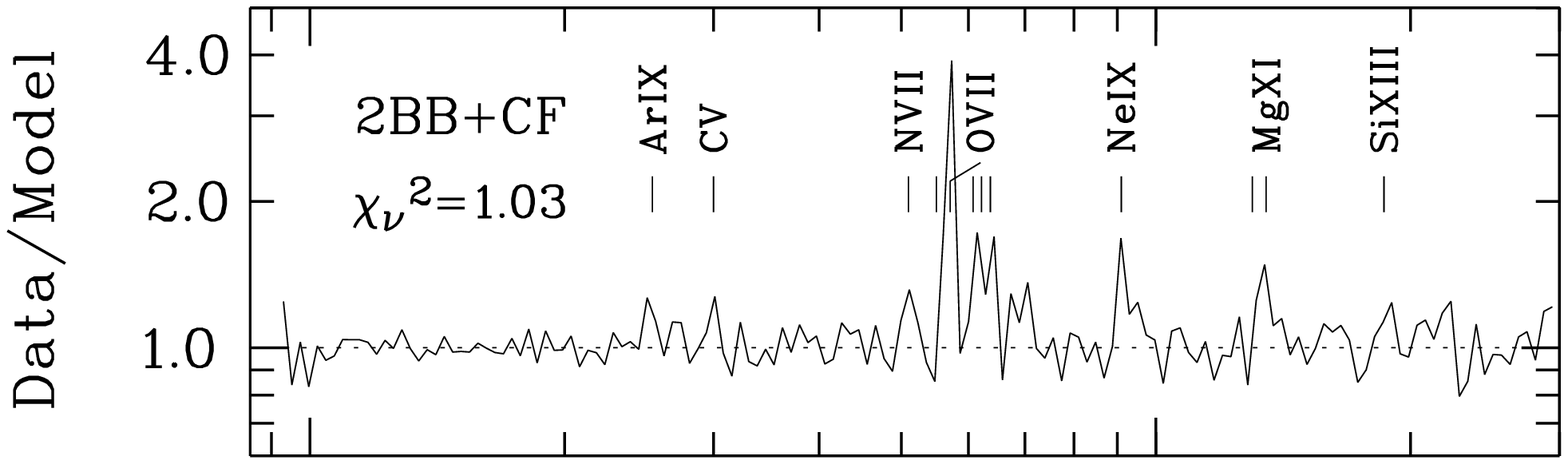}
\includegraphics[width=76.6mm,angle=0,clip]{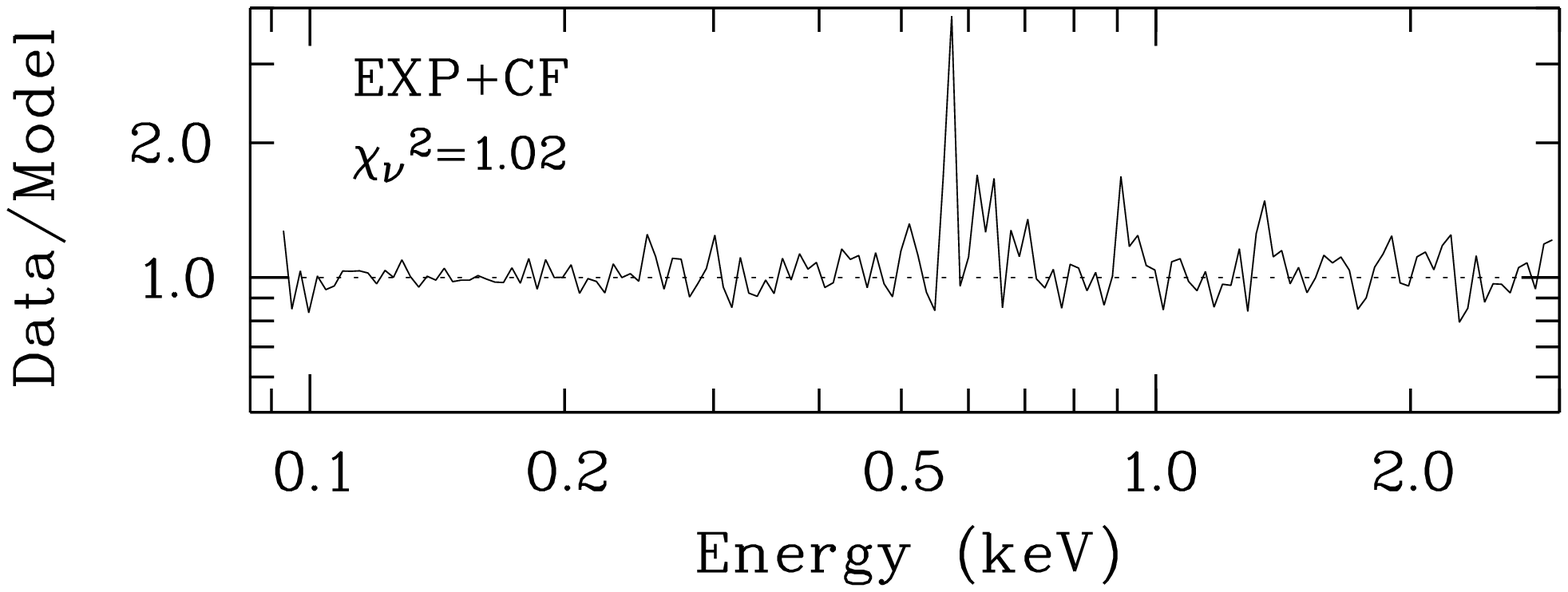}
\end{minipage}
\caption[chart]{\emph{Left:} Mean \chandra\ LETG spectrum of September
  2000 reduced to the first diffraction order (green curve) and fitted
  with the EXP+CF model (upper black solid curve, see text).
  \emph{Right, top:} Residuals for the 1BB+CF model represented by the
  data/model flux ratio. The reduced $\chi^2$ value refers to the
  energy range of 0.10\,--\,0.50 keV. \emph{Center:} Same for the
  2BB+CF model. The twelve spectral bins excluded in the fits are
  indicated by tick marks. \emph{Bottom:} Same for the EXP+CF model
  displayed in the left panel.}
\label{fig:chandra}
\end{figure*}

We analyzed the \chandra\ observation 645 of 30 September 2000, which
was performed with the LETG and the HRC-S as detector and lasted
24\,ks. The data cover the wavelength range 4--120\,\AA\ (0.1--3.0
keV) with a spectral resolution of $\Delta
\lambda\mathrm{(FWHM)}\simeq 0.05$\,\AA. The time-dependent spectra
were analyzed by \citet{burwitzetal02} and only the time-averaged mean
is considered here. The spectrum was re-binned from its original
resolution of 0.05\AA\ to a logarithmic wavelength scale with 100 bins
per decade, which averages over the multitude of faint emission and
absorption features but leaves the major spectral emission lines
intact (Fig.~\ref{fig:chandra}, left panel, solid green curve).
Absolute spectral fluxes were derived using the LETGS effective areas
given in the \chandra\ calibration files CALDB 4.4.7
\footnote{The calibration files can be found at 
 http://cxc.harvard.edu/caldb/ downloads/Release\_notes/CALDB\_v4.4.7.html.
For a general approach see also \citet{beuermannetal06}.}. 
We re-analyzed, in addition, the \rosat\ PSPC observation rp300067 of
12/13 April 1991 \citep{gaensickeetal95,ramsayetal96,beuermannetal08}. 
This spectrum contains 1.3 million counts and has a statistical error
near 0.20\,keV of 0.39\% per 10-eV PSPC channel. Both data sets
provide nearly full orbital phase coverage. On both occasions,
\amher\ was in its normal accretion mode \citep{mazehetal86} with
visual AAVSO magnitudes at orbital maximum of $V\simeq13.3$ (\chandra)
and 13.2 (\rosat), respectively. The PSPC spectrum was analyzed using
the detector response matrix DRMPSPC\_AO1c \citep{beuermann08}.

% Fig. 1

\begin{table*}[t]
\caption{Parameters of spectral fits to the mean September 2000
  \chandra\ LETG and April 1991 \rosat\ PSPC spectra of \amher.  The
  upper index '15eV' refers to spectral components with blackbody
  temperatures k$T>15$\,eV. BB=blackbody model,
  CF=cooling-flow model, EXP=exponential model (Eq.~1), RL='He-like
  emission lines' excluded from the fit to the \chandra\ LETG
  spectrum. }
\label{tab:amher}
\begin{tabular}{l@{\hspace{4mm}}c@{\hspace{2mm}}c@{\hspace{4mm}}cccccc}
\hline \hline \noalign{\smallskip}
 Model & \nh~$^{1)}$  & k$T$ & k$T_0$ & \fbb & $F^\mathrm{15eV}_\mathrm{bb}$ & $f_{930}$ & $f^\mathrm{15eV}_{930}$ &\hspace{-3mm}$\chi^2/\rm{d.o.f.}$ \\[0.3ex] 
   & ($10^{19}$\atoms) & (eV) & (eV)  &  \multicolumn{2}{c}{($10^{-9}$\ergs)} &  \multicolumn{2}{c}{($10^{-13}$\ergsa)} &\\ 
\noalign{\smallskip} \hline \noalign{\smallskip}
\multicolumn{5}{l}{\textit{(a) \chandra\ LETG, September 2000:}}\\[0.5ex]
1\,BB+CF~$^{2)}$  & $4.35\pm 0.10$ & $32.3\pm 0.3$ && \multicolumn{2}{c}{$1.15\pm 0.05$} & \multicolumn{2}{c}{$0.11\pm 0.01$}  & \hspace{-2.5mm}297.1\,/\,63~$^{3)}$ \\ 
2\,BB+CF~$^{2)}$  & $6.00\pm 0.44$ & $24.7\pm 1.9$ && \multicolumn{2}{c}{$2.60\pm 0.60$} &  \multicolumn{2}{c}{$0.50\pm 0.20$} & \hspace{-1mm}63.8\,/\,61~$^{3)}$    \\ 
                  &                & $41.5\pm 2.1$ &&                                    &                                     &                                    \\ 
10\,BB+CF~$^{2)}$ & $6.71\pm 0.00$ &12.5\ldots 57.5 && $5.57\pm 0.85$ & $3.95\pm 0.25$   & $3.00\pm 0.93$ & $1.18\pm 0.18$     & \hspace{-1mm}63.3\,/\,56~$^{3)}$    \\
EXP+CF~$^{2)}$    & $6.57\pm 0.11$ && $3.45\pm 0.04$ & $5.74\pm 0.30$ & $4.14\pm 0.20$   & $4.06\pm 0.25$ & $1.33\pm 0.08$     & \hspace{-1mm}64.2\,/\,63~$^{3)}$    \\[1.0ex]
\multicolumn{6}{l}{\textit{(b) \rosat\ PSPC, April 1991:}}\\[0.5ex]	   
1\,BB+CF+RL~$^{4)}$& $4.25\pm 0.15$ & $30.7\pm 0.3$ && \multicolumn{2}{c}{$1.29\pm 0.06$} & \multicolumn{2}{c}{$0.14\pm 0.01$} & 50.6\,/\,66~$^{5)}$                 \\
EXP+CF+RL~$^{4)}$ & $6.72\pm0.17 $ & 28.5~$^{6)}$   & $3.43\pm 0.06$ & $5.47\pm 0.33$     & $4.22\pm 0.10$ & $3.37\pm 0.29$    & $1.37\pm 0.09$ & 45.5\,/\,65~$^{5)}$\\[0.5ex]
\noalign{\smallskip} \hline \noalign{\smallskip}
\end{tabular}

$^{1)}$ $N_\mathrm{H}$ is for a hydrogen ionization $x_\mathrm{H}\!=\!0.25$. 
$^{2)}$ Fit excludes the helium-like emission lines. 
$^{3)}$ $\chi^2$ and the number of d.o.f refer to the 0.10--0.50\,keV energy interval.
$^{4)}$ RL component included in the fit. 
$^{5)}$ $\chi^2$ and d.o.f. refer to 0.09--2.00\,keV. 
$^{6)}$ Average count rate-weighted blackbody temperature.
\end{table*}

\section{Model assumptions}

Our model includes 
%the usual 
two dominant spectral components, thermal
hard X-rays from the post-shock cooling flow (CF) and soft X-rays
produced by various modes of reprocessing of the energy transported
from the CF into the atmosphere of the white dwarf
\citep{kuijperspringle82,franketal88,gaensickeetal95,koenigetal06}. In
addition, a small component of residual emission lines (RL) exists
that is not explained by the CF model and may be due to
photoionization \citep{mukaietal03,girishetal07}.  We describe the
soft X-ray component by single and multi-temperature variants of the
blackbody model (1BB, 2BB, 10BB, EXP), further explained in the next
Section.

The thermal X-ray emission of the CF is described as optically thin
free-free and bound-free emission for a plasma of solar abundances,
employing Mekal spectra. This model neglects the influence that
cyclotron emission exerts on the temperature distribution and is
appropriate for \amher\ as a low-field polar \citep{fischerbeuermann01}. 
Depending on the specific accretion rate, the shock may be
free-standing above the atmosphere or buried in the photosphere and a
variety of such columns may coexist in the spot at any given time. For
matter of solar composition, the shock temperature is
$T_\mathrm{s}\!=\!32.2\,(M/\mathrm{M_\odot})/(R/10^9\,\mathrm{cm})$\,keV,
or $T_\mathrm{s}\!=\!33.5$\,keV for the white dwarf in \amher\ with
mass $M\!=\!0.78$\,\msun\ and radius $R\!=\!\ten{7.5}{8}$\,cm
\citep{gaensickeetal06}. X-rays emitted from buried columns can escape
only by passing through atmospheric material with column densities
ranging up to
$N_\mathrm{H}\ga10^{25}$\,\atoms\ \citep{beardmoreetal95,ishidaetal97,christian00}.
We include a partial absorber with a column density $N_\mathrm{H,int}$
and a covering fraction $f_\mathrm{c}$, but disregard the reflection
component, which becomes prominent at higher photon energies
\citep{vanteeselingetal96}. This simplified treatment is adequate for
the limited energy range covered by the LETGS, but can not be expected to
provide an accurate measure of the bolometric flux contained in the
hard X-ray component (see Sect.~4.4).

We included photoabsorption by an interstellar column density \nh\ of
partially ionized matter with solar abundances and parts of the metals
condensed into dust. The ionization fractions of hydrogen and helium,
\xh\ and \xhe, are parameters of the fit, with the resulting value of
\nh\ depending on \xh. Most results are quoted for \xh=0.25 and
\xhe=0.40, as derived for near-solar interstellar space
\citep[see][their Sect. 4.2]{beuermannetal06}.
%, but we comment on the \nh(\xh) dependence.
We included the NeVI absorption edges at 78 and 85\AA\ detected
in the EUVE spectrum of \amher\ \citep{paerelsetal96}, which are either of
interstellar or circumbinary origin. The respective optical depths,
$\tau_{78}$ and $\tau_{85}$, are free parameters of the fit.

\section{Results}

The CF component reproduces the 0.6-3.0\,keV part of the LETG spectrum
and accounts for the hydrogen-like emission lines of nitrogen, oxygen,
magnesium, neon, silicon, and sulphur, but largely lacks the
helium-like triplets and some other lines. We excluded, therefore,
twelve of 151 spectral bins from the fit that contain these lines and
represent the RL component (marked by ticks in Fig.~\ref{fig:chandra},
center right panel). The fit yields a covering fraction
$f_\mathrm{c}\!\simeq\!0.70$, implying that 30\% of the CF component
escape unattenuated and 70\% pass through an internal absorber of
$N_\mathrm{H,int}\simeq\ten{3.1}{23}$\,\atoms.  

Interstellar absorption by a column density \nh\ acts on both, the
soft and hard X-ray components. For our standard values \xh=0.25 and
\xhe=0.40, the derived column densities in Table~1 exceed the value of
$\ten{(3.0\pm1.5)}{19}$\,\atoms\ derived from \lya\ absorption in the
FUV spectrum of \amher\ \citep{gaensickeetal95}. The fit value of
\nh\ depends, however, on \xh\ and for our final model the \lya\ value
is reproduced for \xh~$=0.68\pm0.18$. The true value of the degree of
ionization along the line of sight to \amher\ is not well known and we
can not exclude such a high value. We note that \amher\ itself does
not add much to the mean ionization along the path, because its
Stroemgren radius amounts only to a few pc for a mean atomic hydrogen
density of 0.25\,hydrogen atoms\,cm$^{-2}$. Our fits confirm the
absorption edges at 78 and 85\,\AA\ found by \citet{paerelsetal96},
although with smaller optical depths $\tau_{78}\simeq0.05$ and
$\tau_{85}\simeq0.11$.

Obtaining a formal \,$\chi^2$ of the fit to the LETG spectrum requires
an assessment of the errors of the binned spectral fluxes. This is not
straightforward, because faint emission and absorption lines are
spread over the entire wavelength range
\citep{burwitzetal02,trill06}. Since these features are not accounted
for by our model, we defined the wavelength-dependent `error' for the
$\chi^2$ calculation by a polynomial fit to the rms deviations between
data and the best model vs. wavelength, with the twelve spectral bins
defining the RL component excluded. This fixes the residual
$\chi^2_\nu$ of the best-fitting models near unity. The adopted error
varies from 5\% of the first-order flux at 0.1\,keV to 13\% at
3.0\,keV.

\subsection{Modeling the soft X-ray component of \amher}

\emph{1BB+CF model: } The single-blackbody model utterly fails to fit the
low-energy part of the LETG spectrum. Our fit yields an unacceptable
reduced ~$\chi^2_\nu\!=\!4.71$ (Table~1). The residuals, in the form of
the ratio data/model, are displayed in Fig.~\ref{fig:chandra} (top
right panel). The model involves two free parameters for the soft X-ray
component, three parameters for the CF component, and three parameters
describing the interstellar absorption (\nh, $\tau_{78}$ and
$\tau_{85}$, with \xh\ and \xhe\ fixed). The fit parameters and the
derived fluxes are summarized in Table~1.

\emph{2BB+CF model: } Adding a second blackbody (and two free parameters)
significantly improves the fit with best-fit temperatures of 24.7 and
41.5\,eV. The residuals of the fit are displayed in the center right
panel of Fig.~\ref{fig:chandra}.  The fact that the 2BB+CF model fits
the LETG spectrum so well might suggest a dichotomy in the blackbody
temperature of the accretion spot. On theoretical grounds, we consider
it more plausible, however, that the result indicates the presence of
a wider range of blackbody temperatures in the time-averaged spectrum.

% Fig. 2
\begin{figure}[t]
\begin{center}
\includegraphics[width=88.0mm,angle=0,clip]{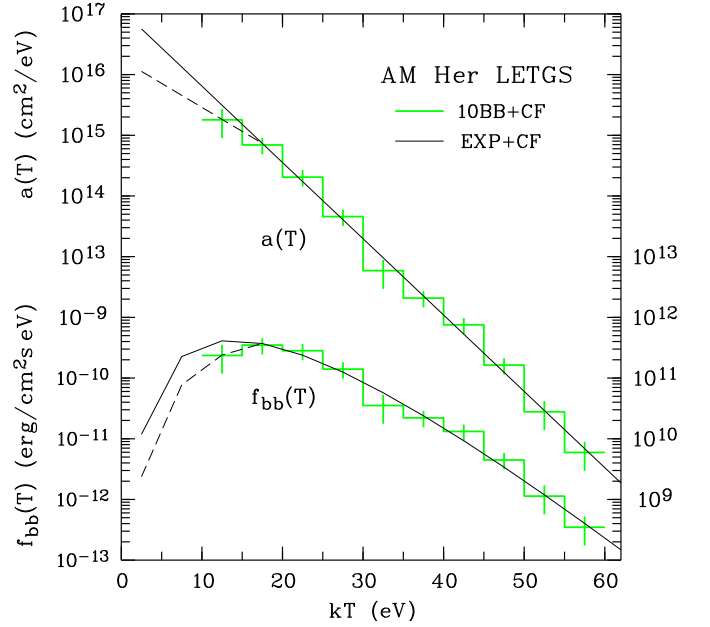}
\caption[chart]{Projected emitting area per eV on the white dwarf in
  \amher\ and integrated blackbody flux per eV at the Earth as a
  function of the blackbody temperature k$T$ for the 10BB+CF
  model. The solid line represents the EXP model of Eq.~1 (see
  text).}
\label{fig:10temp}
\end{center}
\end{figure}

\emph{10BB+CF model: } We study the temperature distribution in more
detail, by introducing a grid of 10 fixed blackbody temperatures,
which covers the interval 10\,--\,60\,eV in 5 \,eV steps with
k$T_\mathrm{n}\!=\!2.5\,(2n\!-\!1)$\,eV and $n\!=\!3$ to 12. The fit
turns out to be stable if a small amount of regularization in the
normalizations is introduced. We left the normalizations for nine of
the ten bins free and prevented a loose-end excursion in the first bin
by tying it to the trend defined by the other normalizations.
Fig.~\ref{fig:10temp} shows the best-fit projected emitting area per
eV, $a(T)$, and the corresponding unabsorbed integrated blackbody flux
per eV, $f_\mathrm{bb}(T)$. The latter peaks in the 15--20\,eV bin and
substantial contributions are expected to exist at still lower
temperatures. The \emph{observed} soft X-ray spectral flux lacks these
low-temperature contributions because of interstellar absorption. Only
a meagre 1.2\% of the integrated observed flux originate from
k$T\!<\!15$\,eV.  The value of ~$\chi^2$ is slightly improved for the
10BB model compared with the 2BB case, but ~$\chi^2_\nu$ deteriorates
from 1.03 to 1.13 (see Table~1), because the individual normalizations
are not independent of each other. The errors in Fig.~\ref{fig:10temp}
are estimates from runs with different levels of regularization of the
first bin. We conclude that the emitting area on the white dwarf
decreases continuously with increasing temperature, although with ten
free parameters the 10BB model is clearly overdetermined.

\emph{EXP+CF model: } For k$T\!>\!15$\,eV, the projected emitting area
$a(T)$ of the 10BB model can be approximated by an exponential
\begin{equation}
a(T) = a_0\,\mathrm{exp}(-T/T_0)
\label{eq:exp}
\end{equation} 
shown as the solid line in in Fig.~\ref{fig:10temp}. For a distance of
AM Her of 80\,pc \citep{thorstensen03,beuermann06}, we find
\mbox{$a_{\,0}\!=\!\ten{1.17}{17}$ cm$^2$eV$^{-1}$} and
k$T_0\!=\!3.46$\,eV. The standard version of our EXP model utilizes
the full temperature range and fits the soft X-ray emission of
\amher\ excellently with only one free parameter besides the
normalization. Hence, the EXP model is mathematically as simple as the
single blackbody. A fit using this model is displayed in
Fig.~\ref{fig:chandra} (left panel and bottom right panel). The
parameters and derived fluxes are given in Table~1.

The EXP model is attractive, because it provides a heuristic
description of the time-averaged mean temperature distribution in the
photospheric accretion spot on the white dwarf in \amher. The
theory of the heating and cooling processes in the individual subcolumns of the
spot is not yet well developed \citep[e.g.][]{koenigetal06} and the
present analysis may help to improve it.

As shown in the Appendix, some key quantities of the EXP model can be
expressed analytically. For the temperature range from k$T\!=\!15$\,eV to
infinity (or $\sim\!100$\,eV) and $a_0$ and k$T_0$ as derived from the
best fit, the projected area of the soft X-ray emitting spot
is $A_{\perp,X}\!=\!\ten{4.8}{15}$\,cm$^2$ at $d\!=\!80$\,pc, corresponding to
0.07\% of the surface area of the white dwarf. For still lower
temperatures, the spot area increases, but can no longer be determined
from the fit to the LETG spectrum alone.

\subsection{The FUV flux of \amher}

In the EXP model, temperatures k$T\!\la\!15$\,eV contribute little to
the soft X-ray flux detected with the LETGS, but produce a sizeable
fraction of the FUV flux. In principle, the run of $a(T)$ at low
temperatures can be obtained from FUV observations, subject, however,
to substantial observational and theoretical
uncertainties. Observationally, the FUV flux is dominated by emission
from the accretion stream, with the Balmer (and presumably the Lyman)
jump strongly in emission. The spot contribution, on the other hand,
is probably characterized by rapid variability and the lack of strong
line emission. Whether it shows the Lyman jump in emission or
absorption remains uncertain
\citep{gaensickeetal95,gaensickeetal98,greeleyetal99,hutchingsetal02,
  koenigetal06}. For simplicity, we retain the multi-temperature
blackbody model.
%For simplicity, we continue to use the blackbody
%assumption to calculate the expected FUV flux for out models.

The mean orbital spectral flux of \amher\ at 930\AA\ was measured with
the shuttle-based Hopkins Ultraviolet Telescope (HUT) in March 1995
\citep{greeleyetal99} and with the Far Ultraviolet Spectroscopic
Explorer (FUSE) in June 2000 \citep{hutchingsetal02,gaensickeetal06}
as $f_{930}\simeq\ten{3.5}{-13}$ and $\ten{2.6}{-13}$\,\ergsa, when
the mean AAVSO visual magnitudes of \amher\ were $V\simeq13.2$ and
13.3, respectively. \citet{greeleyetal99} reported the presence of a
rapidly fluctuating `flare minus non-flare' component (FNF) with a
mean-orbital 930\,\AA\ spectral flux of $\ten{1.3}{-13}$\,\ergsa.
Corrected for reddening, the quoted fluxes become
$f_{930}\simeq\ten{4.0}{-13}$\,\ergsa\ (HUT),
$\ten{3.0}{-13}$\,\ergsa\ (FUSE), and $\ten{1.5}{-13}$\,\ergsa\ (HUT,
FNF). We compare these observed fluxes with the predictions of the
spot models presented in the last Section. We do not correct for the
orbital modulations of the individual components, because the
observations are not simultaneous, the individual high states have
different flux levels, and the blackbody assumption adds to the
uncertainties.

% Fig. 3
\begin{figure}[t]
\includegraphics[height=88mm,bb=71 68 434 621,angle=270,clip]{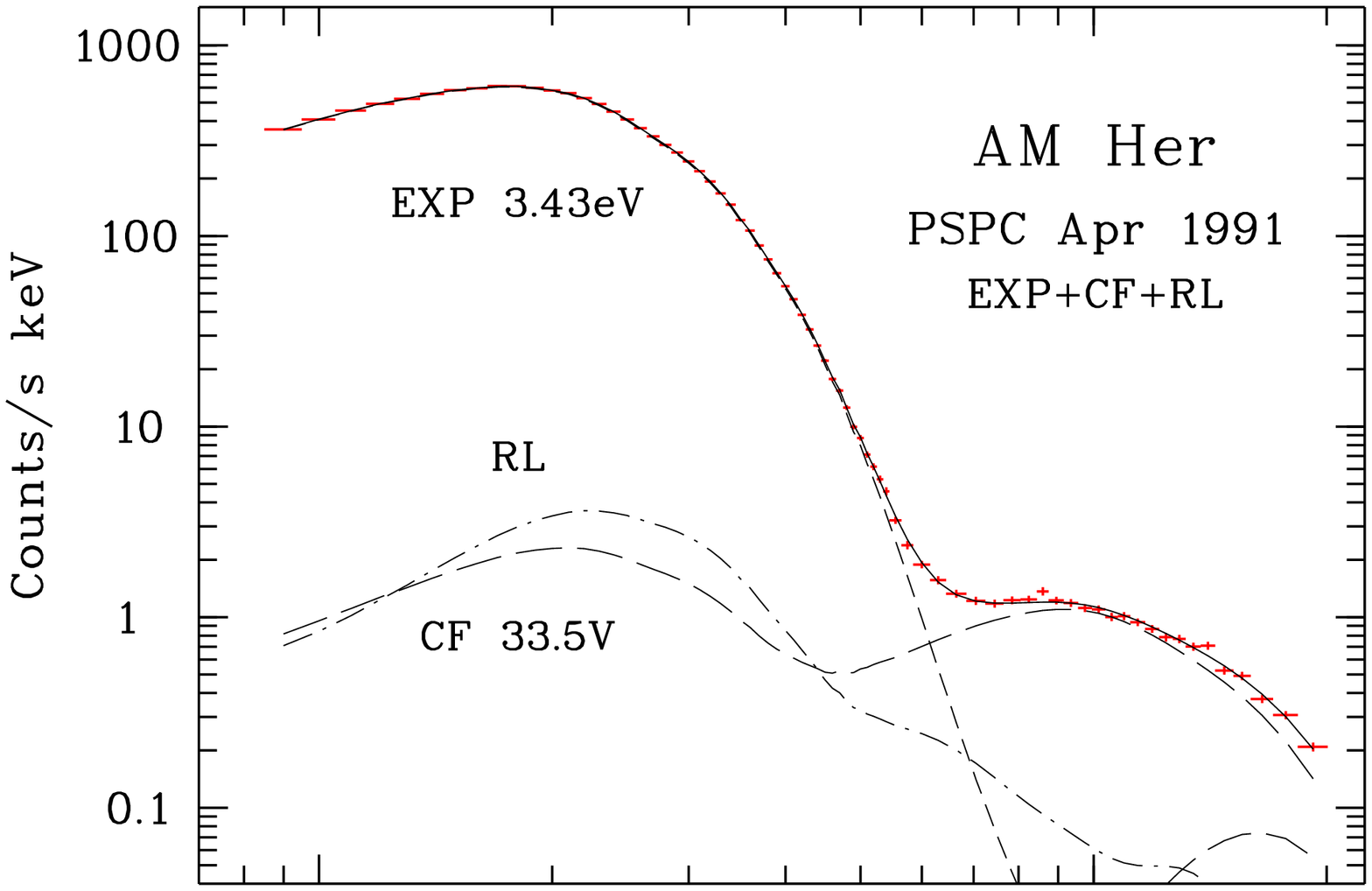}
\includegraphics[height=88mm,bb=306 68 479 621,angle=270,clip]{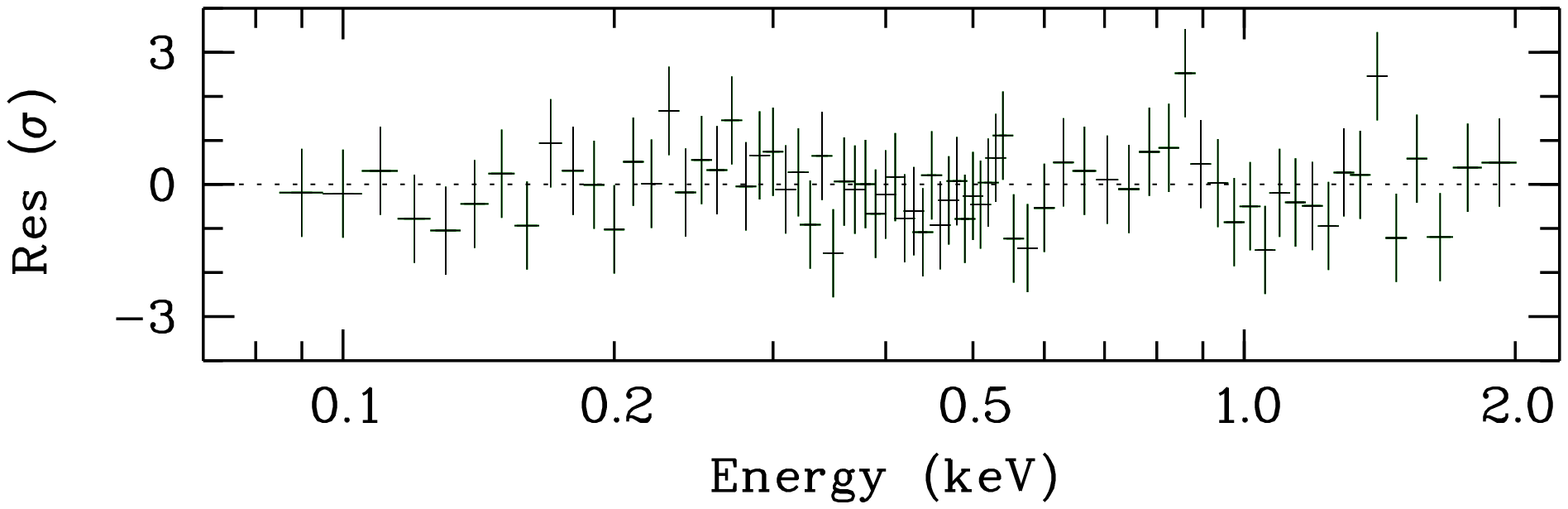}
\caption[chart]{\emph{Top:} 2BB, LIN, and CF components of the September
2000 \chandra\ LETG spectrum of AM Her fitted to the mean April 1991
PSPC spectrum. \emph{Bottom:} Residuals of the fit based on the purely
statistical errors of the PSPC spectrum.}
\label{fig:ampspc}
\end{figure}

The EXP model yields a 930\,\AA\ spectral flux
$f_{930}\!=\!\ten{8.0}{-13}$ \ergsa\ if integrated over all
temperatures and $f_{930}^{15eV}\!=\!\ten{1.35}{-13}$\,\ergsa\ for
\mbox{k$T\!>\!15$\,eV. The} former exceeds the reddening-corrected
total HUT and FUSE 930\,\AA\ fluxes by a moderate factor of two, while
the latter agrees approximately with the lower HUT FNF flux. Taken at
face value, these numbers suggest that the EXP model overestimates the
emitting area at k$T<15$\,eV. Reducing $a(T)$ by a factor of four at
5\,eV (dashed line in Fig.~\ref{fig:10temp}) yields an acceptable
$f_{930}\!=\!\ten{4.06}{-13}$\,\ergsa, while
$f_{930}^{15eV}\!=\!\ten{1.33}{-13}$\,\ergsa\ stays practically
unchanged. These are the numbers included in Table~1. Since the
inherent uncertainties in this comparison may easily amount to a
factor of two, we conclude, that the EXP model or its variant with a
reduced emitting area for k$T\!<\!15$\,eV provide an adequate
description of the FUV flux expected from the entire hot polar cap of
\amher.  The fraction of the flux that arises from k$T\!>\!15$\,eV,
agrees with Greeley's FNF component, suggesting that it represents the
Rayleigh-Jeans part of the spectral flux emitted by the proper
accretion spot. For comparison, the 1BB model fails to predict the FUV
flux by more than an order of magnitude (Table~1).

\subsection{Fitting the \rosat\ PSPC spectrum}
\label{sec:pspcfit}

Finally, we show that the EXP+CF+RL model fits the April 1991
\rosat\ PSPC spectrum, too. We allowed for small corrections to the
nominal calibration of the PSPC as prescribed by \citet{prietoetal96}
and \citet{beuermann08}. There are four free parameters of the fit,
two for the EXP component and two for the normalizations of the
numerically provided RL and total CF components taken from the fit to
the \chandra\ LETG spectrum. The excellent fit is shown in
Fig.~\ref{fig:ampspc} and the derived parameters are listed in
Table~1.  For the PSPC fit, the normalization of the EXP component is
almost identical to that of the LETG, while the CF-component is 24\%
weaker and the RL component 27\% stronger, differences that are easily
accounted for by time variability of the source
\citep{stellaetal86,trill06}. An acceptable fit to the PSPC spectrum
is also obtained for the 1BB+CF+RL model (Table~1), demonstrating that
the higher resolved LETG spectrum is needed to define a superior model
for the soft X-ray emission of \amher.

\subsection{Accretion luminosity and accretion rate}

Soft and hard X-ray emission are the dominant accretion-induced
radiation components in \amher.  Table~2 summarizes their integrated
fluxes, which refer to the bright phase (magnetic phases
$\phi_\mathrm{magn}\!=\!0.2-1.0$) and exceed the mean orbital fluxes
in Table~1 by about 22\%.  The fluxes $F$ are converted to
luminosities $L\!=\!\eta \pi d^2F$, assuming the geometry factors
$\eta$ given in the table and a distance $d\!=\!80$\,pc
\citep{thorstensen03,beuermann06}. The ultraviolet flux from the
heated polar cap of the white dwarf is not included, because it is
created by the downward hard X-ray and cyclotron fluxes
\citep{gaensickeetal95,gaensickeetal98,koenigetal06} and is already
accounted for by the choice of the geometry factors. The factor
$3.3\pi$ for the hard X-rays accounts for the reflection albedo
\citep{vanteeselingetal96}. We adopt here Christian's (2000)
integrated hard X-ray flux. The total accretion-induced luminosity
then is \lacc~$\!=\!\ten{2.5}{33}$\,\erg\ and the implied accretion
rate for a white dwarf with a mass of 0.78\,\msun\ and a radius of
$\ten{7.5}{8}$\,cm \citep{gaensickeetal06} is
\mdot~$\!=\!\ten{2.9}{-10}$\,\msunyr. This result is similar to the
PSPC result of \citet{beuermannetal08}, but is placed here on more
secure grounds, being based on the much better resolved \chandra\ LETG
spectrum. The luminosity emitted as soft X-rays exceeds the sum of
the hard X-ray and the cyclotron luminosities by a factor
$L_\mathrm{soft X}/(L_\mathrm{hard X}+L_\mathrm{cyc})\simeq
4.3\pm2.0$. The conservatively estimated error accounts for a
$\sim30$\% uncertainty in the hard X-ray flux, the remaining error in
the bolometric soft X-ray flux, and variations between the individual
high states in which the measurements were taken. A moderate dominance
of soft X-rays is clearly present and is easily explained by the
concept of 'buried shocks' \citep{kuijperspringle82,franketal88}.

% Table 2
\begin{table}[t]
\caption{Bright-phase high-state bolometric energy fluxes of \amher}
\label{tab:fluxes}

\begin{tabular}{l@{\hspace{4.0mm}}c@{\hspace{3.0mm}}c@{\hspace{3.0mm}}c@{\hspace{3.0mm}}c}
\hline \hline \noalign{\smallskip}
Component & $F$  & $\eta \pi$ & $L$    & Ref.\\           
          &(\ergs)&        & (\erg) &    \\           
\noalign{\smallskip} \hline
\noalign{\smallskip}
Soft X, BB k$T>15$\,eV & \hspace{-1mm}$5.0\times 10^{-9}$& $2\pi$   & $1.9\times 10^{33}$ & 1\\
Hard thermal X, CF     & $3.1\times 10^{-10}$            & $3.3\pi$ & $2.0\times 10^{32}$ & 2\\  
                       & $6.4\times 10^{-10}$            & $3.3\pi$ & $4.0\times 10^{32}$ & 3\\
Cyclotron              & $1.1\times 10^{-10}$            & $2\pi$   & $4.2\times 10^{31}$ & 4\\       
Accretion stream       & $1.7\times 10^{-10}$            & $4\pi$   & $1.3\times 10^{32}$ & 4\\       
                          
\noalign{\smallskip} \hline  \noalign{\smallskip}
\end{tabular}             
                          
References: (1) LETG EXP fit, this work; (2) LETG CF fit, this work,
$N_\mathrm{H,internal}\!=\!\ten{3}{22}$\,\atoms; (3) RXTE, Christian
(2000), $N_\mathrm{H,internal}$ up to $10^{25}$\,\atoms; (4) Optical/IR, Bailey et al. (1988), G\"ansicke et al. (1995).
\end{table}

\section{Conclusions}
\label{sec:con}

We have presented a new model for the soft X-ray emission of polars,
which is superior and physically more plausible than the frequently
employed single-blackbody model. Our model provides an excellent fit
to the high-resolution \chandra\ LETG spectrum of \amher\ and fits the
\rosat\ PSPC spectrum equally well. Without further assumptions, the
model provides also an explanation of the rapidly varying FUV spectral
flux of \amher\ \citep{greeleyetal99}, which may represent the
Rayleigh Jeans tail of the soft X-ray component. On theoretical
grounds, the intermittent heating and subsequent cooling of individual
segments of the accretion spot must lead to a spatially and temporally
averaged temperature distribution in the spot, which extends from
several tems of eV down to the photospheric temperature. Our model
provides a first approximation to this temperature distribution, using
a single-parameter fit to the soft X-ray spectral energy
distribution. The new model is mathematically as simple as the
isothermal single-blackbody model and physically more plausible.

\begin{acknowledgements}  
We thank the anonymous referee for useful comments, which helped to
improve the presentation.
%The observational part of this 
%research was in part funded by the DLR under project number
%50\,OR\,0501.
This research has made use of data obtained from the Chandra Data Archive under ObsId 00645, and software provided by the Chandra X-ray Center (CXC).
\end{acknowledgements}

\appendix

\section{Analytical expressions}

Let the emitting area be $a(T)\!=\!a_0\,exp(-T/T_0)$ and the emitted
spectrum at temperature $T$ be a blackbody. For a temperature
interval of $T_1$ to infinity and $x\!=\!T/T_0$, the
projected area of the spot is
\begin{equation}
A_{\perp} = \Omega d^2= a_{0}\,T_0\int_{T_1}^\infty \mathrm{e}^{-x}\,\mathrm{d}x 
= a_0\,T_0\,\,\mathrm{e}^{-x_1},
\end{equation}
with $\Omega$ the solid angle subtended by the spot and $d$ the
distance. With $C\!=\!a_{0}\,\sigma T_0^5/(\pi d^2)$, where
$\sigma$ is the Stefan-Boltzmann constant, the integrated unabsorbed
blackbody flux at the Earth is
\begin{equation}
F_\mathrm{bb} = C\hspace{-1mm}\int_{T_1}^\infty \hspace{-2.5mm}x^4\mathrm{e}^{-x}\,\mathrm{d}x =
C(x_1^4+5\,x_1^3+12\,x_1^2+24\,x_1+24)\,\mathrm{e}^{-x_1}\!.~~
\end{equation}
The mean spot temperature weighted by the unabsorbed blackbody flux is
\begin{equation}
\langle T_\mathrm{bb}\rangle\,=T_0\left(5+\frac{x_1^5}{x_1^4+5\,x_1^3+12\,x_1^2+24\,x_1+24}\right),
\end{equation}
which assumes its minimal value $\langle T_\mathrm{bb}\rangle\,\!=\!5T_0$ if
the exponential extends to zero temperature. 

\bibliographystyle{aa}

\end{document}